\documentclass[a4paper,fleqn,usenatbib]{mnras}

\usepackage{newtxtext,newtxmath}
\usepackage[T1]{fontenc}
\usepackage{ae,aecompl}

\usepackage{graphicx}	% Including figure files
\usepackage{amsmath}	% Advanced maths commands
\usepackage{amssymb}	% Extra maths symbols

\title[RR Lyrae stars in Crater~II]{Variable Stars in Local Group Galaxies. IV.
RR Lyrae stars in the central regions of the low-density galaxy Crater~II\thanks{This 
article is based on observations made in the Observatorios de Canarias del IAC with
the INT operated on the island of La Palma by the Isaac Newton Group in the
Observatorio del Roque de los Muchachos}}

\author[M. Monelli et al.]{
M. Monelli,$^{1,2}$\thanks{E-mail: monelli@iac.es (MM)}
A.R. Walker,$^{3}$
C.E. Mart\'inez-V\'azquez,$^{3}$
P.B. Stetson,$^{4}$
C. Gallart,$^{1,2}$
\newauthor
E.J. Bernard,$^{5}$
G. Bono,$^{6}$
A.K. Vivas,$^{3}$
G. Andreuzzi,$^{7}$
M. Dall'Ora,$^{8}$
G. Fiorentino,$^{9}$
\newauthor
A. Dorta$^{1,2}$
\\
$^{1}$Instituto de Astrof\'{i}sica de Canarias, Calle Via Lactea, E-38205, La Laguna, Tenerife, Spain\\
$^{2}$Universidad de La Laguna, Dpto. Astrof\'{i}sica, E-38206, La Laguna, Tenerife, Spain  \\
$^{3}$Cerro Tololo Inter-American Observatory, National Optical Astronomy Observatory, Casilla 603, La Serena, Chile\\
$^{4}$Dominion Astrophysical Observatory, Herzberg Institute of Astrophysics, National Research Council, Victoria, British Columbia V9E 2E7, Canada\\
$^{5}$Universit\'{e} C\^{o}te d'Azur, OCA, CNRS, Lagrange, F-06304 Nice, France\\
$^{6}$Departimento di Fisica, Universit\'{a} di Roma Tor Vergata, via della Ricerca Scientifica 1, 00133, Rome, Italy\\
$^{7}$Fundaci\'{o}n Galileo Galilei - INAF, Bre\~{n}a Baja, La Palma, Spain\\
$^{8}$INAF-Osservatorio Astronomica di Capodimonte, salita Moiariello 16, 80131, Napoli, Italy\\
$^{9}$INAF-Osservatorio Astronomica di Bologna, via Ranzani 1,40127, Bologna, Italy\\
}

\date{Accepted XXX. Received YYY; in original form ZZZ}

\pubyear{2018}

\begin{document}
\label{firstpage}
\pagerange{\pageref{firstpage}--\pageref{lastpage}}
\maketitle

\begin{abstract}

We present a search and analysis of variable stars in the recently
discovered Crater~II dwarf galaxy. Based on $B$, $V$, $I$ data collected with
the Isaac Newton Telescope (FoV$\sim$0.44 square degrees) we detected 
37 variable stars, of which 34 are bone-fide RR Lyrae stars 
of Crater~II (28 RRab, 4 RRc, 2 RRd).  We applied the metal-independent 
($V$, $B-V$) Period--Wesenheit 
relation and derived a true distance modulus ($\mu$ = 20.30$\pm$0.08 mag
($\sigma$=0.16 mag). Individual metallicities for RR Lyrae stars were derived
by inversion of the predicted $I$-band Period-Luminosity relation. We find a 
mean metallicity of [Fe/H]=-1.64 and a standard deviation of $\sigma_{[Fe/H]}$
=0.21 dex, compatible with either negligible or vanishing intrinsic metallicity
dispersion. The analysis of the Colour-Magnitude Diagram reveals a 
stark paucity of blue horizontal branch stars, at odds with other 
Galactic dwarfs, and globular clusters with similar metal abundances.

\end{abstract}


%%%%%%%%%%%%%%%%%%%%%%%%%%%%%%%%%%%%%%%%%%%%%%%%%%

%%%%%%%%%%%%%%%%% BODY OF PAPER %%%%%%%%%%%%%%%%%%

\section{Introduction}

The last few years have revealed a completely new picture in our understanding
of the galactic neighborhood  with the discovery of  $\sim20$ dwarf galaxies
surrounding the Milky Way, doubling the number of satellites known in 2014 
\citep{bechtol15,koposov15b,drlicawagner15a,laevens15a,laevens15b,kim15a,kim15b,
kim15d,martin15,torrealba16a,torrealba16b,drlicawagner16,torrealba18,koposov18a}.
Standing out among the new
discoveries, the Crater II galaxy \citep{torrealba16a}  is not an ultra-faint
compact satellite, but it is instead the fourth more extended galaxy around the Milky
Way, following the two Magellanic Clouds and the Sagittarius dwarf spheroidal
(dSph) galaxy. Crater II is, however, significantly fainter than those large
galaxies, having an absolute magnitude of $M_{V} = -8.2$ mag
\citep{torrealba16a}. With a surface brightness of only 31 mag/arcsec$^2$, 
Crater II is nearly invisible in a sea of field stars and  lies close to the limits
of detectability of current dwarf satellite searches \citep{koposov08b}.   No
other Milky Way satellite has similar properties of brightness and size as
Crater II although the plot of absolute magnitude vs. radius given by
\citet[][their Figure 6]{torrealba16a}  shows that Andromeda XIX
\citep{mcconnachie08b, cusano13} has similar surface brightness with slightly
greater brightness and size. 

The classical dSph galaxies surrounding the Milky Way have proven to be complex
objects with varied, and different, star formation histories 
\citep[SFH, e.g.][]{mateo98,tolstoy09}.  No two dSph galaxies in our neighborhood are alike and the
reason(s) why they present such a variety of chemical enrichment and SFH is 
not well understood today. Study of the nature of 
these systems is fundamental for interpreting their role in the hierarchical
process of formation of large galaxies such as our own Milky Way. Crater II
presents then a unique opportunity to investigate the stellar population and SFH
in a galaxy unlike any other presently known among the Milky Way satellite
galaxies.  

The Color-Magnitude Diagram (CMD) of Crater II in its discovery paper
\citep{torrealba16a} shows a strong red giant branch (RGB) and horizontal
branch (HB). Comparison with isochrones suggests that age of 10 Gyr and 
metallicity of [Fe/H]=--1.7 represent the data reasonably well. However,
the depth of such CMD, limited to $\sim$0.5 mag below the HB, is inadequate 
to answer the critical question of the extent of the SFH of Crater II.  The
Carina dSph galaxy, for example, also presents a narrow RGB but
deep CMDs clearly show a complex SFH with multiple main sequence turnoffs (MSTO)
and sub-giant branches \citep{io}.

%%%%%%%%%%%%%%%%%%%%%%%%%%%%%%%%%%%%%%%%%%%%%%%%%%%  FIG 1 %%%%%%%%%%%%%%%%%%%%%%%%%%%%%%%%%%%%%%%%%%5
    \begin{figure}
  \vspace{-2.5cm}
  \hspace{-1cm}
   \includegraphics[width=0.48\textwidth]{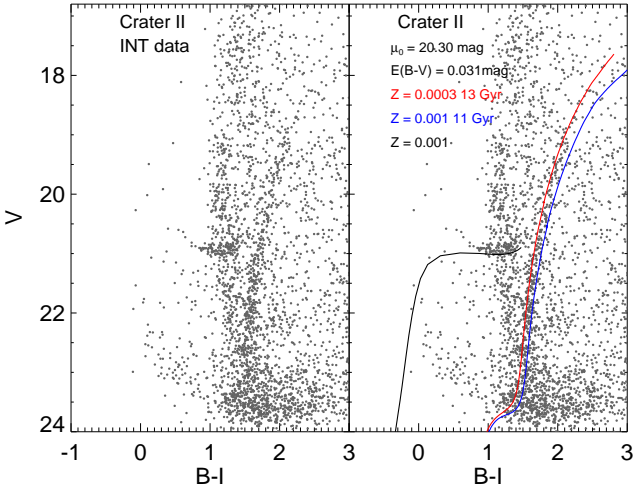}
  \caption{($V$, $B-I$) colour-magnitude of the central region of Crater~II. The bright portion
  of the RGB and the red HB are the most prominent features, together with a heavy contamination
  of field stars. The right panels present the same CMD with superimposed old (13 Gyr 
  and 10 Gyr) and metal-poor (Z=0.0003, 0.001) isochrones from the BaSTI library.
  A ZAHB for Z=0.001 is also plotted, shifted by 0.05 mag \citep{cassisi07}. 
  These theoretical lines were
  shifted according with a distance modulus of 20.30 mag (see \S~\ref{sec:distance} 
  for details) and a reddening of $E(B-V) = 0.027$ mag \citep{schlafly11}.
}
     \label{fig:cmd}
  \end{figure}
%%%%%%%%%%%%%%%%%%%%%%%%%%%%%%%%%%%%%%%%%%%%%%%%%%%%%%%%%%%%%%%%%%%%%%%%%%%%%%%%%%%%%%%%%%%%%%%%

Spectroscopy by \citet{caldwell17a} of 62 stars nominated as members selected from a
sample close to the RGB in the CMD show that Crater II has extremely cold
dynamics, even though it is still dark matter dominated with mass-light ratio 
$53^{+15}_{-11}$ $M_{\odot} /L_{V, \odot}$.   For these stars they also
determine a mean metallicity [Fe/H] = --2.0 $\pm$ 0.1 dex with a resolved
dispersion of $\sigma_{[Fe/H]} = 0.22^{+0.04}_{-0.03}$ dex.  The photometric
metallicity found from isochrone fitting by \citet{torrealba16a} is some 0.3 dex more
metal rich.
%%%%%%%%%%%%%%%%%%%%%%%%%%%%%%%%%%%%%%%%%%%%%%%%%%%  FIG 2 %%%%%%%%%%%%%%%%%%%%%%%%%%%%%%%%%%%%%%%%%%5
    \begin{figure*}
  \includegraphics[scale=0.8]{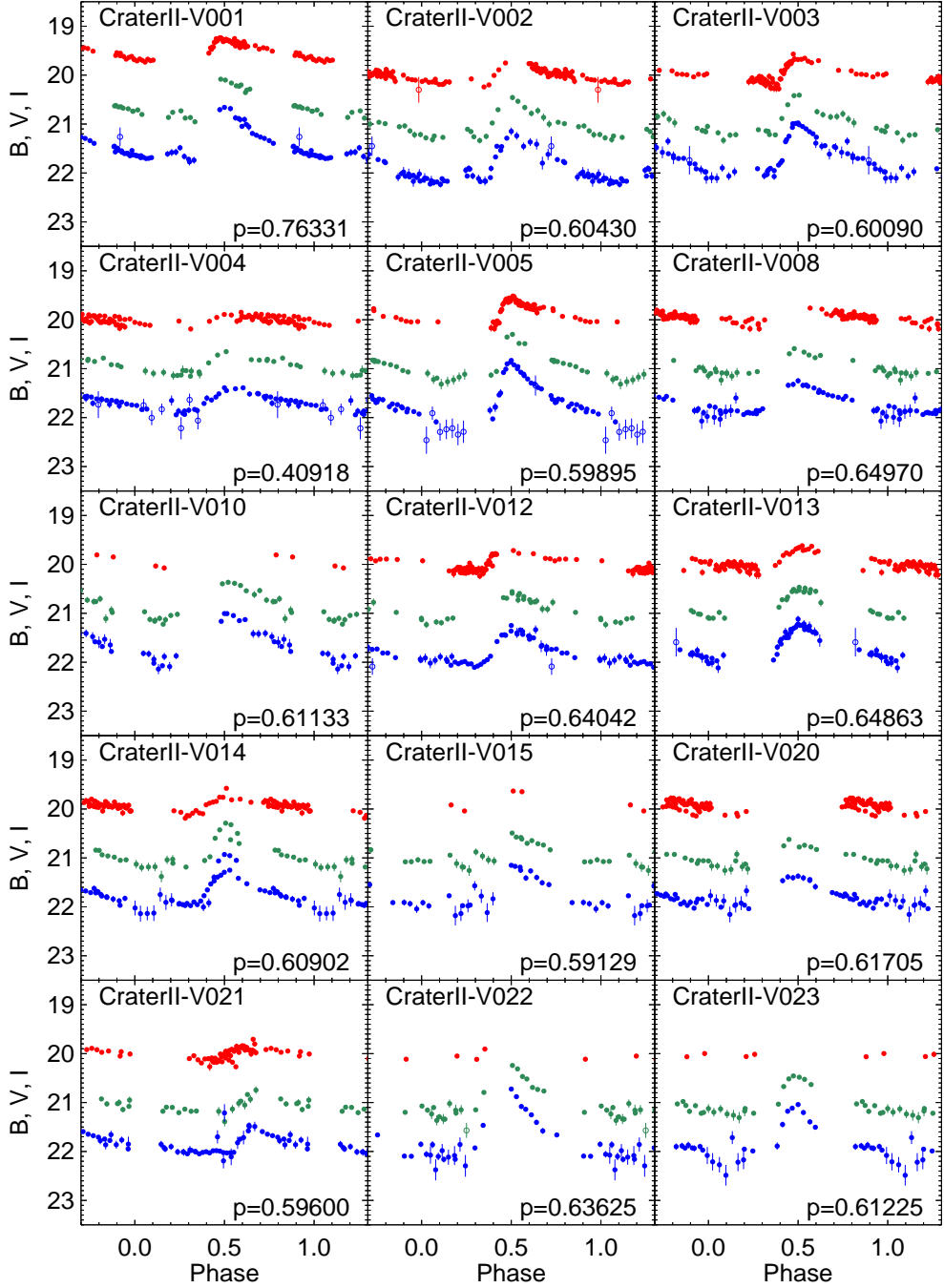}
  \caption{$B$ (blue), $V$ (green), and $I$ (red) light curves for variable stars in Crater~II.
  Star names (upper-left corner) are sorted for increasing right ascension. Periods 
  (lower-right corner) are given in days. Open symbols (not used in the estimation of 
  the pulsation properties) show the data for which the uncertainties are larger than 
  3-$\sigma$ above the mean error of a given star.   Note that $B$ and $I$ light curves 
  were shifted by +0.4 and --0.4 mag, respectively, to facilitate 
  the visual inspection of all three together.}
     \label{fig:lcv1}
  \end{figure*}
%%%%%%%%%%%%%%%%%%%%%%%%%%%%%%%%%%%%%%%%%%%%%%%%%%%%%%%%%%%%%%%%%%%%%%%%%%%%%%%%%%%%%%%%%%%%%%%%55 

%%%%%%%%%%%%%%%%%%%%%%%%%%%%%%%%%%%%%%%%%%%%%%%%%%%  FIG 2b %%%%%%%%%%%%%%%%%%%%%%%%%%%%%%%%%%%%%%%%%%5
\renewcommand{\thefigure}{\arabic{figure} (Cont.)}
\addtocounter{figure}{-1}
\begin{figure*}
  \includegraphics[scale=0.8]{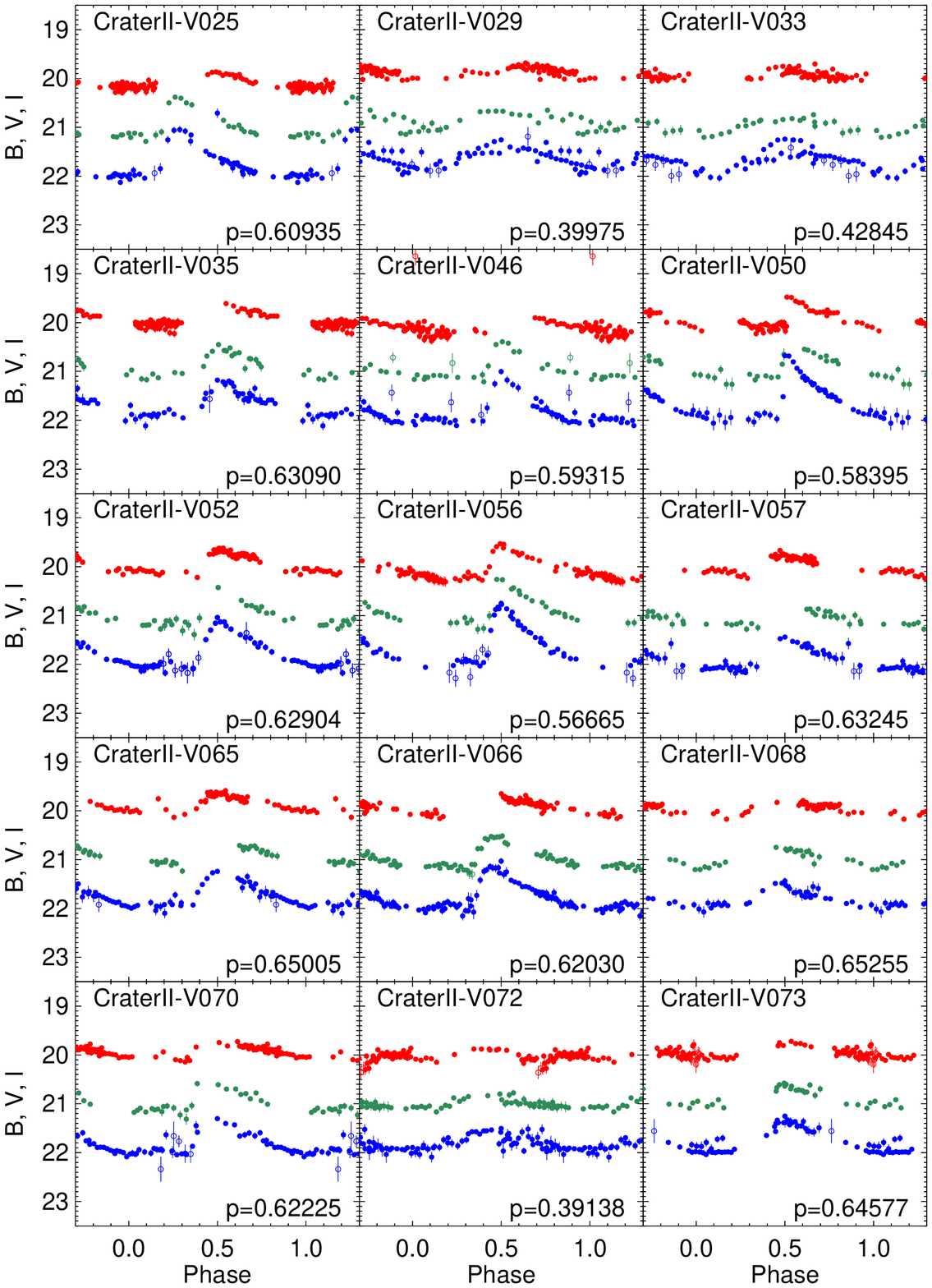}
  \caption{Continuation of Figure \ref{fig:lcv1}.}
     \label{fig:lcv2}
  \end{figure*}
\renewcommand{\thefigure}{\arabic{figure}}
%%%%%%%%%%%%%%%%%%%%%%%%%%%%%%%%%%%%%%%%%%%%%%%%%%%%%%%%%%%%%%%%%%%%%%%%%%%%%%%%%%%%%%%%%%%%%%%%55 

%%%%%%%%%%%%%%%%%%%%%%%%%%%%%%%%%%%%%%%%%%%%%%%%%%%  FIG 2c %%%%%%%%%%%%%%%%%%%%%%%%%%%%%%%%%%%%%%%%%%5
\renewcommand{\thefigure}{\arabic{figure} (Cont.)}
\addtocounter{figure}{-1}
\begin{figure*}
  \includegraphics[scale=0.8]{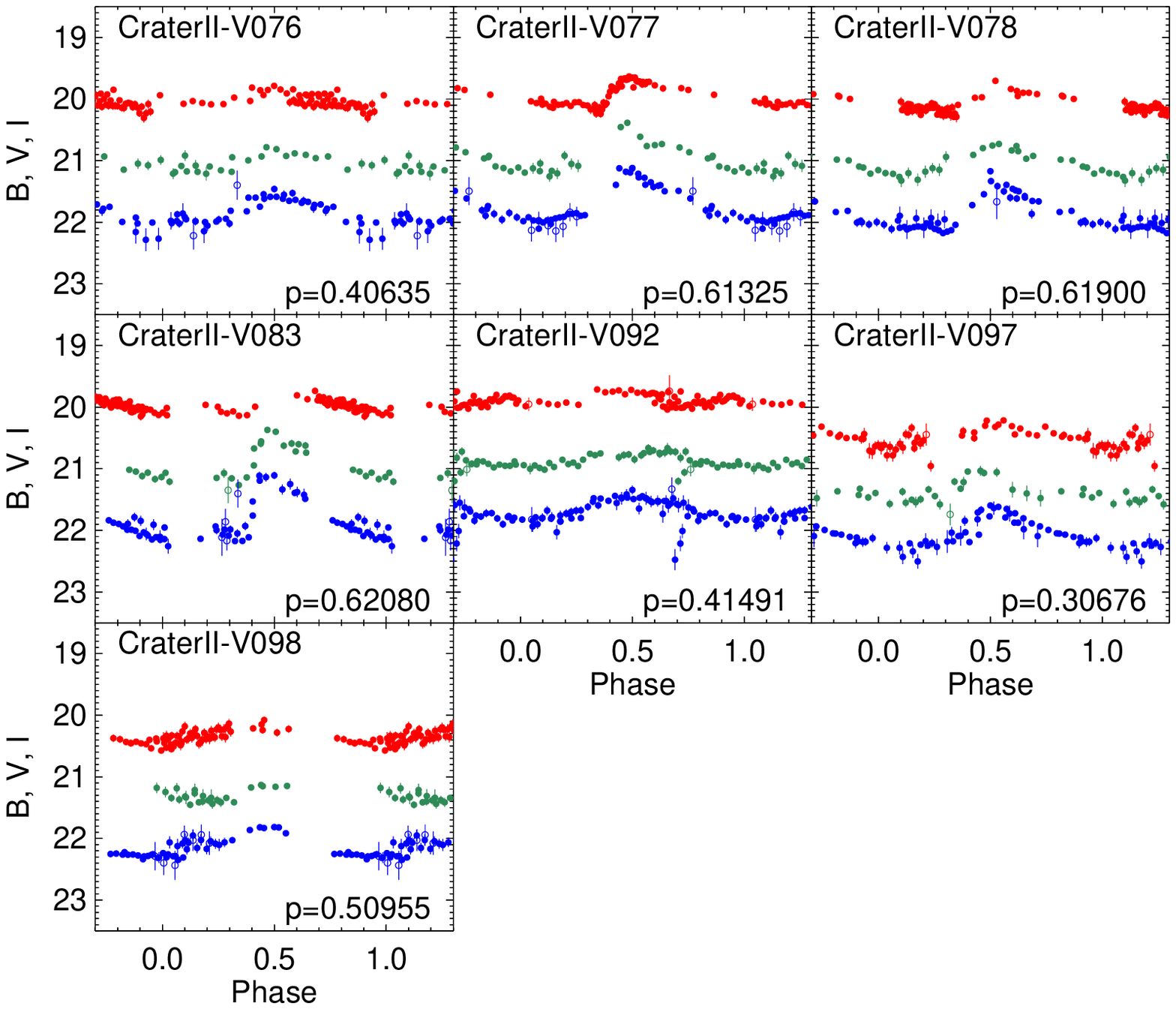}
  \vspace{-8cm}
  \caption{Continuation of Figure \ref{fig:lcv1}.}
     \label{fig:lcv3}
  \end{figure*}
\renewcommand{\thefigure}{\arabic{figure}}
%%%%%%%%%%%%%%%%%%%%%%%%%%%%%%%%%%%%%%%%%%%%%%%%%%%%%%%%%%%%%%%%%%%%%%%%%%%%%%%%%%%%%%%%%%%%%%%%55 

Variable stars can also be used for studying the content and structure of
stellar systems. The presence of RR Lyrae (RRL) stars, for example, is an unequivocal
sign of an old stellar population \citep[$>$ 10 Gyr][]{walker89}, and in addition allows an
independent determination of distance \citep{marconi15,degrijs17d}.  Their period and
mean magnitude distributions can shed light on the metallicity distribution and
even provide hints on the SFH of the oldest populations \citep{martinezvazquez16a}.  
Anomalous Cepheids, which are common (but not very numerous) in dSph galaxies but not
in galactic globular clusters, are usually
interpreted as belonging to an intermediate-age population
\citep{fiorentino12b}, although this interpretation needs care because of
primordial binaries being an alternate production channel \citep{bono97e}.

In this paper we present a search for variable stars in the field  of
Crater~II,  independent from the very recent work by \citet{joo18}.
\S \ref{sec:data} presents a summary of the data and the data
reduction strategy. \S \ref{sec:cmd} presents the CMD and discusses the stellar
populations in Crater~II. \S \ref{sec:vars} discusses the search for variable stars
and the properties of RRL stars. In \S \ref{sec:distance} we discuss
the distance to Crater~II, based on the Period-Wesenheit relation, while in
\S \ref{sec:metal} we derive the RRL stars metallicity distribution. \S 
\ref{sec:discussion} provides a discussion of the results and final
remarks.

\section{Observations and data reduction}\label{sec:data}

Observations were collected with the Isaac Newton Telescope (INT) located in the Observatorio
Roque de los Muchachos (ORM) in La Palma (Spain) using the Wide Field Camera (WFC, 
FoV=33$\arcmin$.8 $\times$ 33$\arcmin$.8), under two programmes approved by the 
Spanish Time Allocation Committee: the Large Programme C143 and the service program 
SST2017-380 (P.I. Monelli). The observations have been realized in three different 
runs, of four, two, and three nights, respectively (see Table 
1 for more details). The area covered by these data is 37$\arcmin\times$62$\arcmin$,
approximately centred on the galaxy, but only the central 0.44 sq.deg. have a robust calibration.
Therefore, this data set covers much less than half the entire body of Crater II \citep[r$_h$=31.2\arcmin,][]{torrealba16a}.

%%%%%%%%%%%%%%%%%%%%%%%%%%%%%% TAB 1 %%%%%%%%%%%%%%%%%%%%%%%%%%%%%
 \begin{table}
 \begin{center}
 \caption{Log of observations}
% %\scriptsize {
 \begin{tabular}{cccccc}
 \hline
 \textit{Run Dates} & \textit{Telescope} & \textit{Detector} & \textit{$B$} & \textit{$V$} & \textit{$I$}  \\
  \hline
% %----------------------------------------------------------------------                                                                   
    2016 May 29-31   & INT &  WFC &  13  &   9  & 10    \\
    2017 Jan 27-28   & INT &  WFC &  21  &  39  & 20    \\
    2017 Apr 03      & INT &  WFC &  41  & ---  & 38    \\
 \hline
 \end{tabular}
 \end{center} 
% %}
 \label{tab:tab01}
 \end{table}

%%%%%%%%%%%%%%%%%%%%%%%%%%%%%%%%%%%%%%%%%%%%%%%%%%%%%%%%%%%

The data have been collected over a 311 d baseline. Given
its declination (Dec $\sim$ --18.5$^\circ$), Crater~II is visible with airmass
$<$1.7 for only about 4 hr per night at the ORM, which implies that
long time-series in a single night could not be obtained. 
Nevertheless, the median (maximum) number of phase
points per star is of 49 (75), 25 (48), and 62 (68) for the $B$, $V$, and $I$
bands, respectively. Therefore, despite the time sampling being not optimal, 
for most of the stars identified as possible
variables we have been able to derive light curve parameters adequate for our
purpose here, though we do not discard that some periods estimated in 
this work may be affected by aliasing problems.

The photometric analysis was carried out by PBS using software, procedures, 
and standard stars that have been extensively described in the literature
{\citep[e.g.,][]{stetson00,stetson05a}.
%PASP 112, 925; 2005 PASP 117 563).

%%%%%%%%%%%%%%%%%%%%%%%%%%%%%%%%%%%%%%%%%%%%%%%%%%%  FIG 3 %%%%%%%%%%%%%%%%%%%%%%%%%%%%%%%%%%%%%%%%%%5
    \begin{figure}
  \vspace{-2.5cm}
   \includegraphics[width=0.55\textwidth]{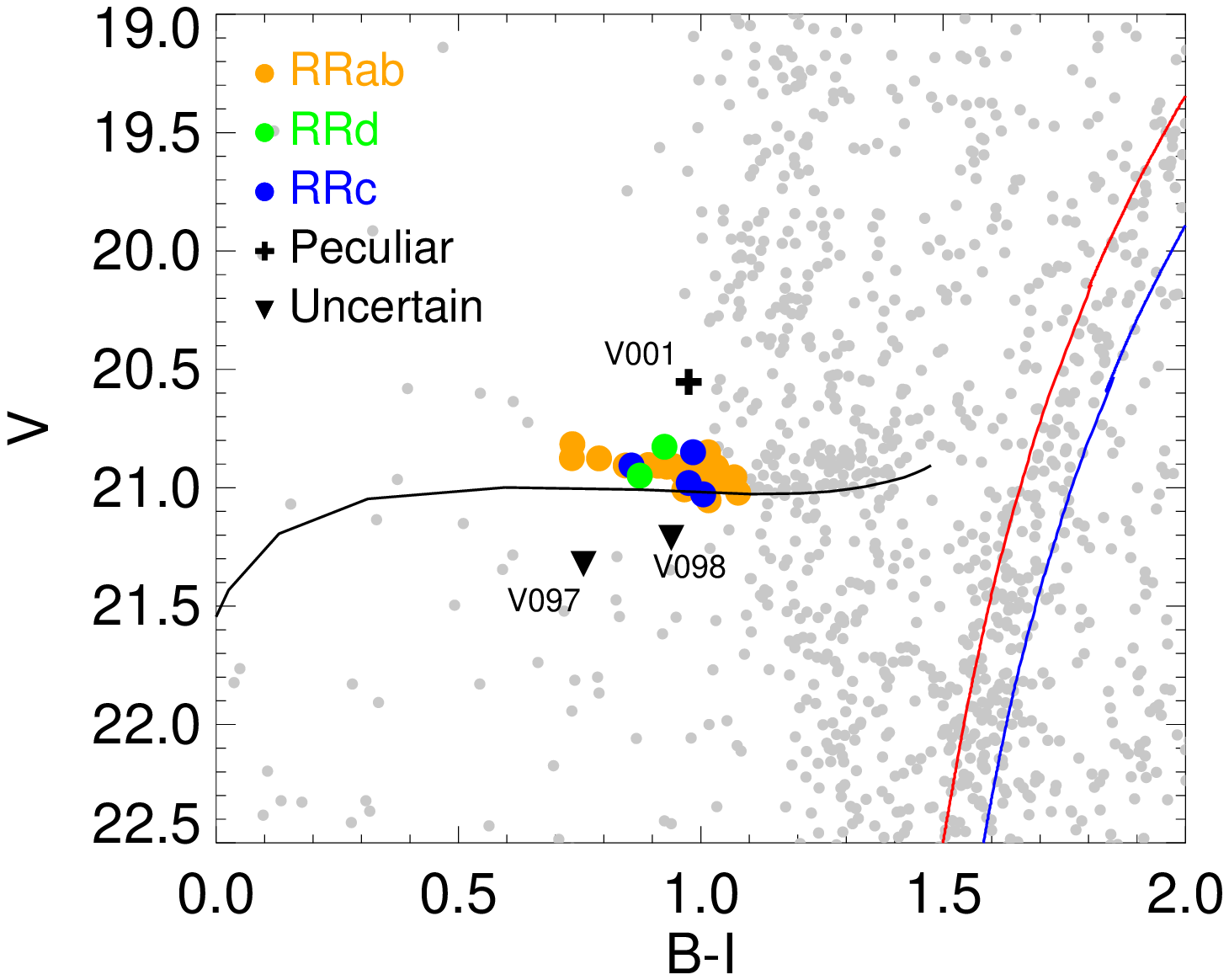}
  \caption{Zoom of the ($V$, $B-I$) colour-magnitude of Crater~II centered on the HB location.
  The discovered RR Lyrae stars are shown with filled symbols, splitting RRab 
  (orange), RRc (blue) and RRd (green) type stars. The same isochrones and ZAHB as in Figure \ref{fig:cmd}
  are shown.}
     \label{fig:cmdvar}
  \end{figure}
%%%%%%%%%%%%%%%%%%%%%%%%%%%%%%%%%%%%%%%%%%%%%%%%%%%%%%%%%%%%%%%%%%%%%%%%%%%%%%%%%%%%%%%%%%%%%%%%
%%%%%%%%%%%%%%%%%%%%%%%%%%%%%%%%%%%%%%%%%%%%%%%%%%%  FIG 4 %%%%%%%%%%%%%%%%%%%%%%%%%%%%%%%%%%%%%%%%%%
    \begin{figure}
    \hspace{-1.8cm}
  \includegraphics[scale=0.60]{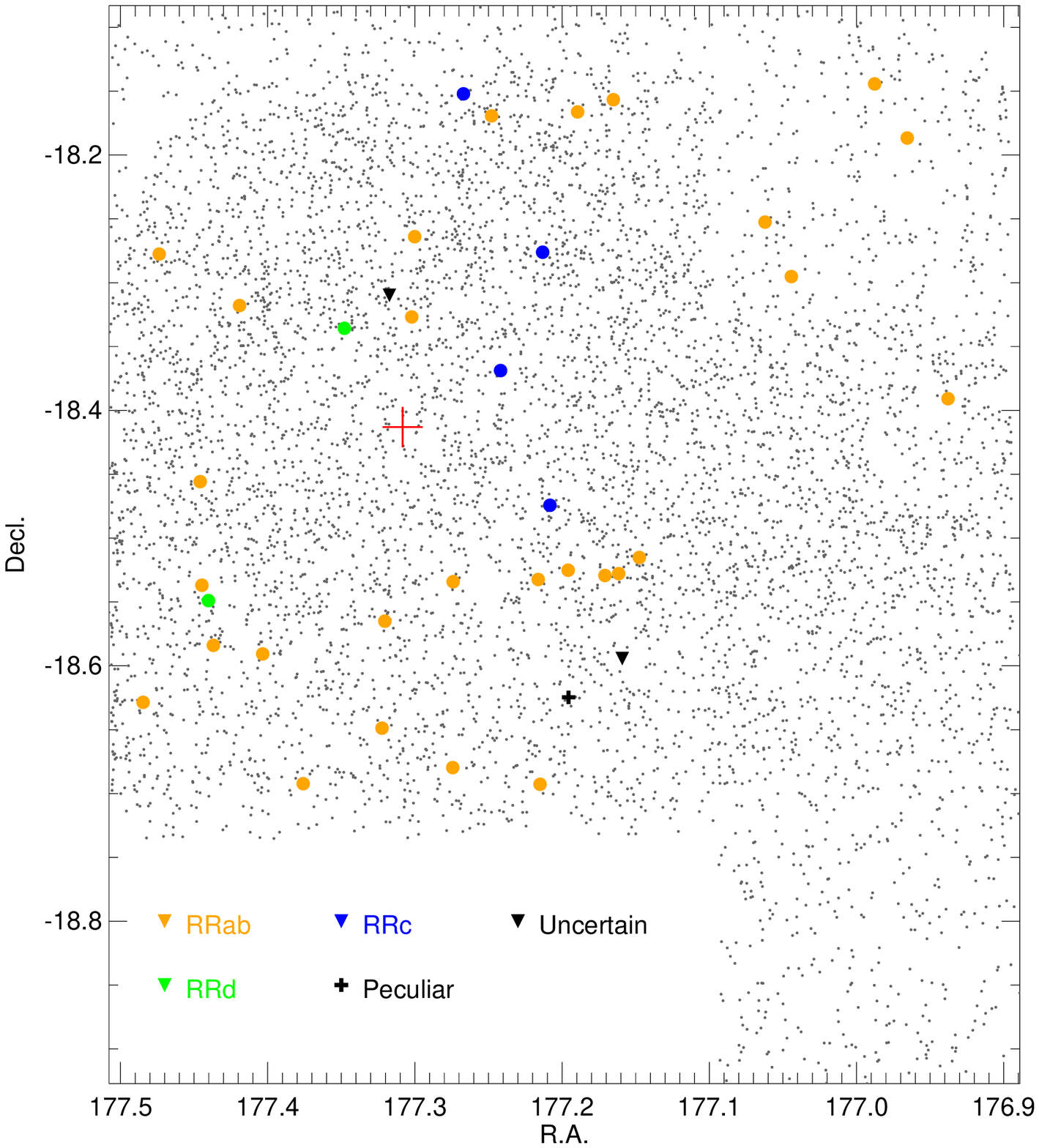}
  \caption{RR Lyrae variable stars in the field of Crater II. The colour code is the same as in
  Figure \ref{fig:cmdvar}.
  Black symbols indicate the position of the peculiar and uncertain stars. The nominal center of the
  galaxy (red cross) is taken from \citet{torrealba16a}.}
     \label{fig:radec}
  \end{figure}
%%%%%%%%%%%%%%%%%%%%%%%%%%%%%%%%%%%%%%%%%%%%%%%%%%%%%%%%%%%%%%%%%%%%%%%%%%%%%%%%%%%%%%%%%%%%%%%%55 

%%%%%%%%%%%%%%%%%%%%%%%%%%%%%%%%%%%%%%%%%%%%%%%%%%%  FIG 5 %%%%%%%%%%%%%%%%%%%%%%%%%%%%%%%%%%%%%%%%%%5
    \begin{figure}
  \includegraphics[width=13cm]{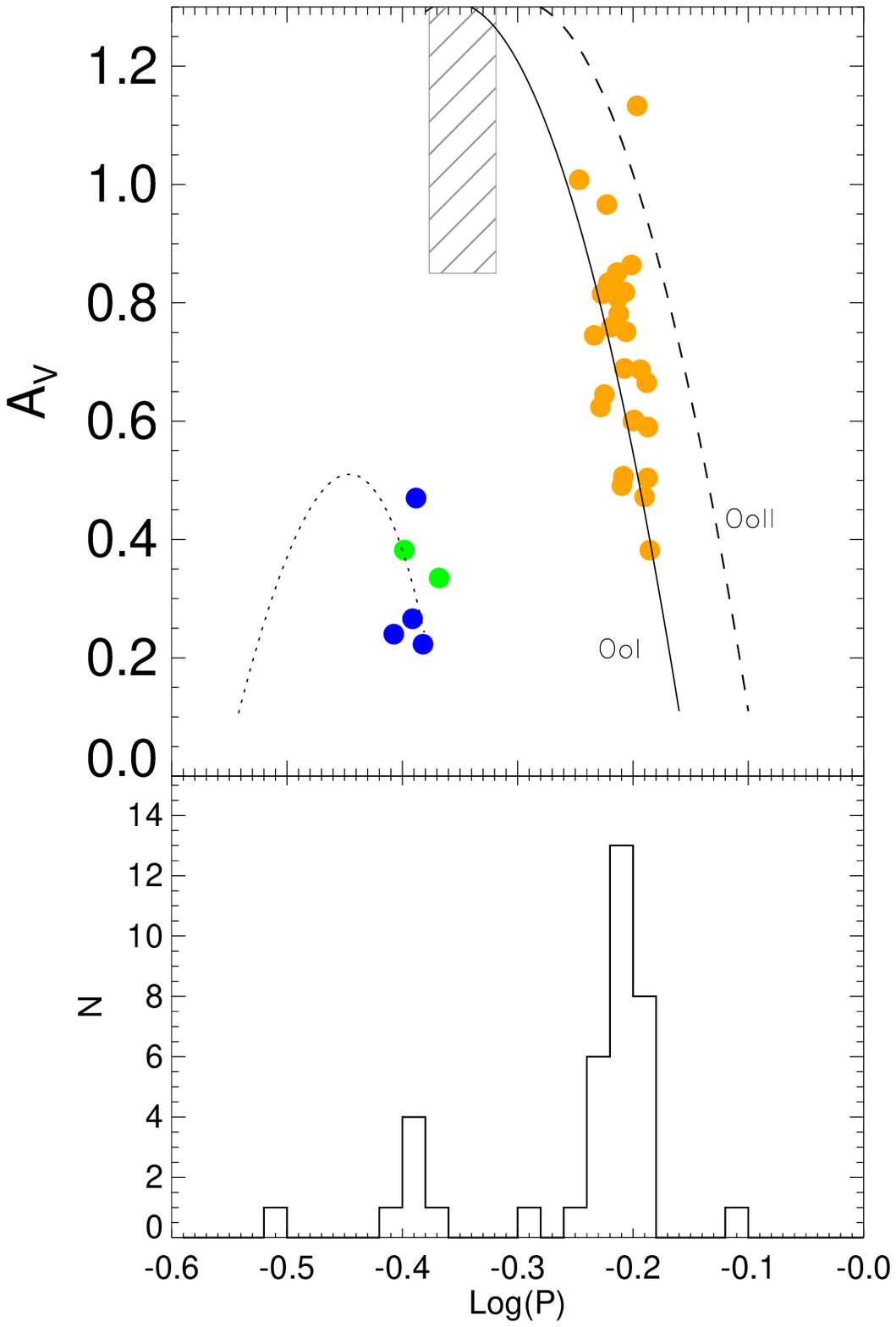}
  \caption{{\it Top -} Period-Amplitude Bailey diagram for the 32 RRL stars in the central
  region of Crater~II. We overplotted the Oosterhoff sequences for fundamental pulsators from 
  \citet[solid and dashed line for OoI and OoII, respectively][]{cacciari05} and for first overtone
  from \citet{kunder13c}. {\it Bottom -} Period distribution for the RRL stars plotted in the top panel.}
     \label{fig:bailey}
  \end{figure}
%%%%%%%%%%%%%%%%%%%%%%%%%%%%%%%%%%%%%%%%%%%%%%%%%%%%%%%%%%%%%%%%%%%%%%%%%%%%%%%%%%%%%%%%%%%%%%%%
%%%%%%%%%%%%%%%%%%%%%%%%%%%%%%%%%%%%%%%%%%%%%%%%%%%  FIG 6 %%%%%%%%%%%%%%%%%%%%%%%%%%%%%%%%%%%%%%%%%%
\begin{center}    
 \begin{figure}
  \includegraphics[width=8.5cm]{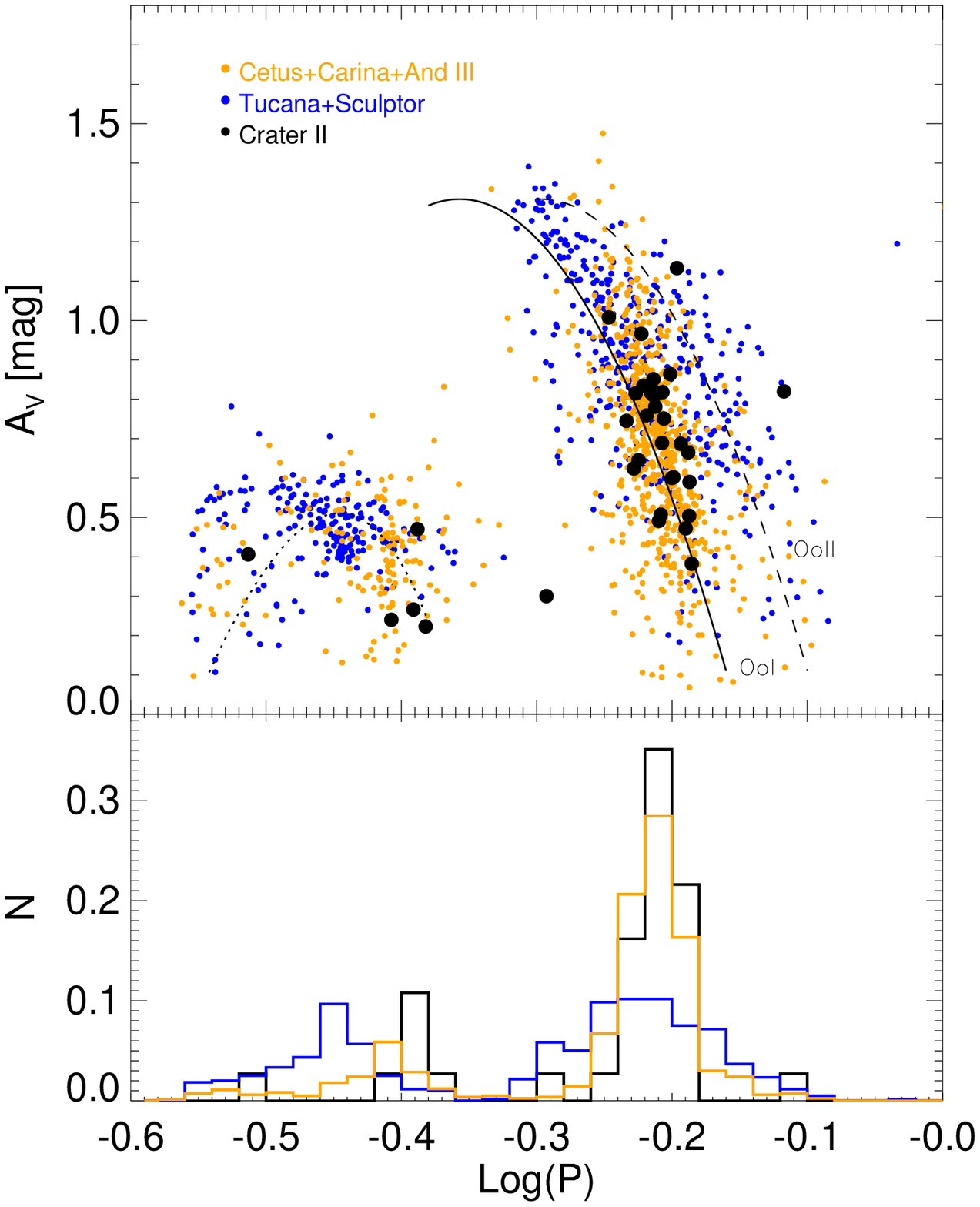}
  \caption{Comparison of the Bailey diagram of different galaxies. Orange symbols show the  
   position of star in Carina, And~III, and Cetus, which present a steeper relation, similar
   to that of Crater~II (black), with respect of Sculptor and Tucana (blue points).}
     \label{fig:bailey_cfr}
 \end{figure}
\end{center}
%%%%%%%%%%%%%%%%%%%%%%%%%%%%%%%%%%%%%%%%%%%%%%%%%%%%%%%%%%%%%%%%%%%%%%%%%%%%%%%%%%%%%%%%%%%%%%%%

%%%%%%%%%%%%%%%%%%%%%%%%%%%%%% TAB 2 %%%%%%%%%%%%%%%%%%%%%%%%%%%%%
 \begin{table*}
 \begin{center}
 \caption{Variable Stars Properties}
 \begin{tabular}{ccccccccccccl}
 \hline
% %-------------------------------------------------------------------------------------
 {\it Name} & {\it R.A.}       & {\it Dec.} & {\it Period}  &  {\it m$_{B}$} & {\it A$_{B}$} & {\it m$_{V}$} & {\it A$_{V}$} & {\it m$_{I}$} & {\it A$_{I}$} & {\it type} & P$_{Joo}$& Notes\\
{\it }     & {\it hr min sec} & {\it $\degr$ $\arcmin$ $\arcsec$} & {\it d}    & {\it } & {\it mag}     & {\it mag}     & {\it mag}     & {\it mag}     & {\it mag}     & {\it mag} &  {\it d}   &  \\
  \hline
% %----------------------------------------------------------------------                                                                   
V001   &   11:48:46.94   &  -18:37:28.8    &   0.7633058   &  20.925   &   0.984    &   20.553    &  0.820  &   19.950   &   0.503   &  \it{Peculiar} &  0.7633     & \\ 
V002   &   11:48:59.49   &  -18:10: 9.8    &   0.6043024   &  21.432   &   0.913    &   20.966    &  0.759  &   20.406   &   0.546   &  RRab    &    0.6053  &  \\ 
V003   &   11:48:38.77   &  -18:31:40.2    &   0.6009024   &  21.217   &   1.019    &   20.920    &  0.834  &   20.349   &   0.515   &  RRab    &    0.6009  &  \\ 
V004   &   11:48:51.17   &  -18:16:34.1    &   0.4091809   &  21.260   &   0.477    &   20.906    &  0.470  &   20.403   &   0.218   &  RRc     &    0.4189  &  \\ 
V005   &   11:48:51.89   &  -18:31:56.7    &   0.5989511   &  21.227   &   1.229    &   20.904    &  0.966  &   20.335   &   0.575   &  RRab    &    0.5989  &  \\ 
V008   &   11:48:47.01   &  -18:31:30.5    &   0.6497012   &  21.217   &   0.679    &   20.907    &  0.504  &   20.305   &   0.372   &  RRab    &    0.6496  &   a\\ 
V010   &   11:47:51.72   &  -18:11:12.1    &   0.6113311   &  21.228   &   1.068    &   20.850    &  0.851  &   20.213   &   0.745   &  RRab    &    0.6212  &   b   \\ 
V012   &   11:48:41.01   &  -18:31:45.4    &   0.6404212   &  21.375   &   0.766    &   20.968    &  0.687  &   20.340   &   0.492   &  RRab    &    0.6415  &   c    \\ 
V013   &   11:48:51.59   &  -18:41:34.0    &   0.6486312   &  21.314   &   0.851    &   20.884    &  0.665  &   20.301   &   0.430   &  RRab    &    0.6474  &  \\ 
V014   &   11:48:39.64   &  -18:09:24.1    &   0.6090211   &  21.310   &   0.735    &   20.906    &  0.816  &   20.329   &   0.390   &  RRab    &    0.6081  &   c  \\ 
V015   &   11:47:57.04   &  -18:08:39.6    &   0.5912911   &  21.289   &   0.917    &   20.904    &  0.624  &   20.280   &   0.424   &  RRab    &    0.6378  &   b; d ; e  \\ 
V020   &   11:48:35.41   &  -18:30:54.6    &   0.6170524   &  21.370   &   0.673    &   20.956    &  0.491  &   20.301   &   0.315   &  RRab    &    0.6201  &   f; a	   \\ 
V021   &   11:49:05.76   &  -18:32:02.9    &   0.5960023   &  21.432   &   0.581    &   21.053    &  0.645  &   20.416   &   0.562   &  RRab    &     0.641  &     \\ 
V022   &   11:48:14.91   &  -18:15:09.1    &   0.6362525   &  21.096   &   1.603    &   20.816    &  1.133  &   20.361   &   0.435   &  RRab    &    0.6001  &   b; e	      \\ 
V023   &   11:48:10.63   &  -18:17:42.7    &   0.6122524   &  21.323   &   1.055    &   20.944    &  0.808  &   20.448   &   0.071   &  RRab    &    0.6147  &   b; e	      \\ 
V025   &   11:49:05.83   &  -18:40:46.9    &   0.6093524   &  21.310   &   1.080    &   20.909    &  0.826  &   20.366   &   0.516   &  RRab    &    0.6083  &   a \\ 
V029   &   11:49:23.49   &  -18:20:08.5    &   0.3997510   &  21.204   &   0.320    &   20.827    &  0.382  &   20.279   &   0.172   &  RRd     &    0.4212  &   \\ 
V033   &   11:49:45.65   &  -18:32:56.2    &   0.4284515   &  21.219   &   0.603    &   20.949    &  0.335  &   20.345   &   0.264   &  RRd     &    0.4174  &   \\ 
V035   &   11:49:56.30   &  -18:37:43.0    &   0.6309025   &  21.260   &   0.714    &   20.879    &  0.600  &   20.255   &   0.444   &  RRab    &    0.6309  &       \\ 
V046   &   11:49:36.81   &  -18:35:26.6    &   0.5931521   &  21.258   &   1.153    &   20.906    &  0.815  &   20.413   &   0.537   &  RRab    &    0.6155  &   g; a\\ 
V050   &   11:49:46.72   &  -18:32:12.9    &   0.5839523   &  21.070   &   1.466    &   20.875    &  0.745  &   20.336   &   0.625   &  RRab    &    0.5877  &   g  \\ 
V052   &   11:49:40.58   &  -18:19:04.4    &   0.6290411   &  21.319   &   0.996    &   20.937    &  0.864  &   20.357   &   0.543   &  RRab    &    0.6280  & \\
V056   &   11:49:44.85   &  -18:35:02.3    &   0.5666522   &  21.184   &   1.264    &   20.876    &  1.008  &   20.394   &   0.746   &  RRab    &    0.5658  &      \\ 
V057   &   11:49:53.68   &  -18:16:39.4    &   0.6324525   &  21.460   &   0.726    &   21.021    &  0.602  &   20.383   &   0.411   &  RRab    &    0.6314  &   g  \\ 
V065   &   11:47:45.07   &  -18:23:27.0    &   0.6500526   &  21.296   &   0.748    &   20.891    &  0.590  &   20.293   &   0.454   &  RRab    &    0.6513  &   g    \\ 
V066   &   11:49:47.00   &  -18:27:20.8    &   0.6203024   &  21.304   &   0.875    &   20.925    &  0.689  &   20.333   &   0.374   &  RRab    &    0.6268  &       \\ 
V068   &   11:49:12.51   &  -18:19:36.5    &   0.6525526   &  21.379   &   0.486    &   20.967    &  0.382  &   20.330   &   0.317   &  RRab    &    0.6478  &     \\ 
V070   &   11:49:12.03   &  -18:15:50.3    &   0.6222525   &  21.342   &   0.904    &   20.912    &  0.751  &   20.311   &   0.514   &  RRab    &    0.6245  &   h  \\ 
V072   &   11:48:49.98   &  -18:28:27.5    &   0.3913809   &  21.347   &   0.379    &   20.980    &  0.240  &   20.372   &   0.143   &  RRc     &    0.6553  &   d     \\ 
V073   &   11:49:30.21   &  -18:41:32.3    &   0.6457655   &  21.357   &   0.720    &   20.910    &  0.472  &   20.335   &   0.343   &  RRab    &    0.6493  &       \\ 
V076   &   11:49:04.08   &  -18:09:07.6    &   0.4063514   &  21.406   &   0.413    &   21.028    &  0.266  &   20.401   &   0.231   &  RRc     &    0.6516  &   d; e\\ 
V077   &   11:49:17.36   &  -18:38:55.8    &   0.6132524   &  21.266   &   0.941    &   20.911    &  0.781  &   20.336   &   0.470   &  RRab    &    0.6132  &     \\ 
V078   &   11:49:16.90   &  -18:33:54.2    &   0.6190025   &  21.403   &   0.865    &   21.007    &  0.507  &   20.437   &   0.447   &  RRab    &    0.6029  &   c     \\ 
V083   &   11:48:45.47   &  -18:09:58.5    &   0.6208012   &  21.382   &   1.197    &   20.916    &  0.818  &   20.350   &   0.518   &  RRab    &    0.6229  & \\ 
V092   &   11:48:58.03   &  -18:22:07.7    &   0.4149109   &  21.247   &   0.372    &   20.850    &  0.223  &   20.263   &   0.173   &  RRc     &    0.7084  &   f  \\ 
V097   &   11:49:16.15   &  -18:18:35.4    &   0.3067604   &  21.625   &   0.649    &   21.322    &  0.406  &   20.867   &   0.364   &  uncertain  & 0.2347	 &  d	 \\ 
V098   &   11:48:38.19   &  -18:35:39.7    &   0.5095519   &  21.664   &   0.454    &   21.212    &  0.300  &   20.725   &   0.261   &  uncertain  &  ---    &   d    \\ 
 \hline
 \multicolumn{12}{l}{Notes:} \\
 \multicolumn{12}{l}{{\it a - }lack of maximum in I band} \\
 \multicolumn{12}{l}{{\it b - }only few points/phase in I band} \\
 \multicolumn{12}{l}{{\it c - }candidate Blazhko} \\
 \multicolumn{12}{l}{{\it d - }noisy light curves} \\
 \multicolumn{12}{l}{{\it e - }many possible periods} \\
 \multicolumn{12}{l}{{\it f - }noisy light curve in I band} \\
 \multicolumn{12}{l}{{\it g - }lack of maximum in V band} \\
 \multicolumn{12}{l}{{\it h - }lack of maximum} \\
 \hline
 \end{tabular}
 \end{center} 
 \label{tab:rrl}
 \end{table*}

\section{Colour-magnitude diagram} \label{sec:cmd}

Figure \ref{fig:cmd} (left panel) presents the ($V$, $B-I$) CMD of the region around
the center of Crater~II. The photometry
reaches $V\sim$23.8 mag, and the plot shows a prominent 
contamination by the Galactic field, at color redder than $B-I\sim$1.0 mag.
Nevertheless, the RGB and predominantly red HB
are clearly visible in the CMD. To guide the eye (right panel), we superimposed two
$\alpha$-enhanched isochrones from the BaSTI \citep{pietrinferni04,pietrinferni06} database. We
assumed the distance estimated in \S \ref{sec:distance}, and the
red and blue lines show the expected location of a population of (Z, age in Gyr) =
(0.0003, 13) and (0.001, 11), respectively. The black line shows the zero age
HB for Z = 0.001, which provides a nice lower envelope for the
distribution of stars in the red part of the HB and for the RRL stars.

The comparison with isochrones clearly shows that these data are too shallow to
detect stars at the turn-off (TO). Nevertheless, a sequence of blue objects
($B-I < 1$, $V > 22$ mag) is well defined between the HB and the TO. Most likely
it is populated by Blue Stragglers, as commonly found in all dSph galaxies
\citep{mapelli07,mapelli09,monelli12a,santana13}. However, based on the present
data we cannot robustly constrain the youngest age at the TO, so we cannot exclude
the present of a small intermediate age component.
Interestingly, the HB of Crater~II is mostly red, with very few, if any, stars
bluer than the instability strip.

\section{Variable stars in Crater~II} \label{sec:vars}

	\subsection{Search and classification} \label{sec:search}
    
The search for candidate variable stars has been realized using the method
introduced by \citet{welch93} and as further developed by \citet{stetson96b}. The
pulsational properties were estimated following \citet{bernard09}. In
particular, a first guess of the period is derived through Fourier analysis of
the time series \citep{horne86}, and then refined by simultaneous visual
inspection of the light curves in all the available filters. Amplitudes are then
estimated by fitting to a set of light-curve templates \citep{layden99}.

    \subsection{RR Lyrae Stars} \label{sec:rrl}

We identified 58 candidate variable stars, of which 37 are confirmed pulsators.
34 of them are RRL stars. Table \ref{tab:rrl} presents the full catalogue, 
including name\footnote{We assumed the naming convention by \citet{joo18}.}, coordinates 
(right ascension and declination), period, classification, followed by 
intensity-weighted mean magnitude and amplitude in the $B$, $V$, and $I$ pass-bands, 
respectively. The last two columns show the period estimate by \citet{joo18} and 
notes on individual stars.} One variable, V098, was not reported in 
\citet{joo18}. We note that two of 
the stars (V023 and V056) had already been reported in the first release of PANSTARR
(\#204928 and \#204866), though they were not associated to Crater~II \citep{sesar17b}.
Time series photometry for the
variable stars detected in this work is listed in the Appendix. Figure 
\ref{fig:lcv1}-\ref{fig:lcv3} present the $B$ (blue), $V$ (green), and $I$ 
(red) light curves of all the discovered RRL variable stars. 

After the submission of this work, \citet{joo18} published an independent
investigation on the variable stars content of Crater~II, based on KMTNet data 
covering 3$^{\circ}\times$3$^{\circ}$ in $B$ and $V$. All variable stars in the 
overlapping area are in common between 
the two photometries except two: based on our data we cannot confirm the variability 
of V093, while we identify a new variable not listed in the \citet{joo18} catalogue
(V098). The comparison of period estimates discloses a good agreement for most of the stars:
the period difference is smaller than 0.05 days for 32 out of 36 stars (88\%). 
Four stars presents larger period differences, between 0.05 and 0.29 d, which is
likely due to aliasing. We detected a small shift in the mean magnitude.
The derived mean magnitudes for our full sample are 
$<B>$=21.30 $\pm$ 0.09 ($\sigma$=0.09) mag, $<V>$=20.92 $\pm$ 0.07 ($\sigma$=0.05) mag, 
and $<I>$=20.34 $\pm$ 0.06 ($\sigma$= 0.05) mag, while both the $<B>$ and $<V>$ of the stars
from \citet{joo18} in common with ours are $\sim$0.03 mag fainter.

Figure \ref{fig:cmdvar} shows a zoom of the CMD in the region of the HB. The RRL
stars are shown as large symbols, with different colours indicating different
pulsation modes: blue, green and orange correspond respectively to the 4 first-overtone (RRc), 2
double-mode (RRd) and the 28 fundamental-mode (RRab) RRL stars discovered in
this work. Three stars are well separated in magnitude from the bulk of the RRL stars. 
V097 and V098 are 0.4 and 0.3 mag fainter than the HB respectively. V098 has a 
period near 0.5d and an RRc-like light curve, with a large phase gap. Additionally, 
inspecting its image shows it may be superimposed on a faint background spiral galaxy.  V097 has a 
typical RRc period.  V001 is 0.5 mag brighter than the HB, with periods and 
colors typical for an RRL star. It looks too faint to be an Anomalous Cepheid (AC), 
which are typically at least 1 mag brighter than the HB, and its period is too
short (P=0.763306 d) to be a BL Her star \citep{dicriscienzo07}. Stars in this position of the 
HB are known in other systems (Carina: \citealt{coppola15}; Sculptor: 
\citealt{martinezvazquez16b}) and may be stars in the final stage of core helium-burning 
evolution evolving rapidly towards the AGB. Following the nomenclature introduced in
in \citet{martinezvazquez16b} for stars with similar behaviour as this one, we classified 
V001 as "peculiar" variable star.  We will not include these three 
stars in the subsequent analysis. 

Figure \ref{fig:radec} presents the spatial distribution of RRL stars. The colour
code is the same as in Figure \ref{fig:cmd}. The red cross marks the position of
the center of Crater~II according to \citet{torrealba16a}. Interestingly, RRL
stars seem to be spread over the full surveyed field, with no obvious
concentration towards the central regions. 

Figure \ref{fig:bailey} shows the period-amplitude (Bailey) diagram (top panel)
and the period distribution (bottom) for the full sample of RRL stars. Lines in
the top panel present a comparison with the Oosterhoff I (Oo-I,
solid) and II (Oo-II, dashed) loci, as defined for RRab type stars in globular
clusters \citep{cacciari05}, and for RRc type \citep[dotted,][]{kunder13c}. The
mean period $<P_{RRab}>$=0.617 d, which would suggest that Crater~II is somewhat more
similar to an Oo-II system, though it is at the edge of the so-called
Oosterhoff gap typically populated by dSph galaxies.
Nevertheless, the top panel of Figure \ref{fig:bailey}
clearly shows that RRab are distributed closer to the Oo-I line, though they
seem to follow a steeper relation. This
translates into a period distribution (bottom panel) which is quite narrow
around the peak. Therefore, the lack of RRab with shortest period and largest amplitude
mimics an Oo-II type
if only the mean period is considered, but clearly the location of stars in the
period-amplitude plane is closer to that of a Oo-I system.
Many other low-mass, metal-poor systems are characterized by similar
period distributions, lacking the short-period tail (e.g. Bootes:
\citealt{siegel06}; And~XI, And~XIII: \citealt{yang12}; And~XIX
\citealt{cusano13}; And~III \citealt{martinezvazquez17}).

Figure \ref{fig:bailey} suggests that the distribution of Crater~II stars in the 
period-amplitude plane follow a distribution steeper than the Oo-I line. This is
similar to what observed in other systems such Carina  \citep{coppola15}, And III
\citep{martinezvazquez17}, and Cetus \citep{monelli12b}. This is illustrated in 
Figure \ref{fig:bailey_cfr}, which compares the Baily diagram (top) of these three
galaxies (orange points) with that of Crater II (black dots). Blue dots show RRL 
stars in Tucana \citep{bernard09} and Sculptor \citep[][their {\itshape clean} 
sample]{martinezvazquez16b}. The plot shows that for amplitude close to
1 mag the orange points tends to split from the clusters'Oo-I line. The normalized
period distribution (bottom panel) of Crater~II RRab stars is remarkably
simialr to that of Carina+Cetus+And~III, while Sculptor and Tucana
present a more extended tail, especially at short period end.

\section{Distance to Crater~II}\label{sec:distance}

RRL stars are primary, fundamental distance indicators \citep{carretta00b,cacciari03}. 
The well-known Metallicity-Luminosity relation \citep{sandage81}
has been extensively used in the literature, applied to globular clusters
\citep{arellanoferro08,greco09,mcnamara11,dicriscienzo11} and nearby dwarf galaxies
\citep[e.g.][]{stetson14b,yang14,martinezvazquez16a,cusano16,cusano17}. 
Nevertheless, this relation is subject to many sources of uncertainties 
\citep[linearity, evolutionary effects, reddening, ][]{bono03b}, though the
most important one comes from the fact that neither the zero point nor the 
slope have been definitively calibrated \citep{degrijs17d}. This translates
into a difference of up to $\sim$0.2 mag according to the adopted calibration
\citep{chaboyer99,bono03a,clementini03}.

Alternatively, period-luminosity relations \citep{longmore86,bono01d} in the near
infrared present important advantages, such as the significantly milder
dependency on the reddening and evolutionary effects. The exhaustive theoretical
work by \citet{marconi15}
discusses in detail the use of Period-Luminosity-Metallicity ($PLZ$) relations
of RRL stars to derive the  distance of the host stellar system. In particular,
the Wesenheit functions \citep{madore82} are particularly useful because, by
construction, they are reddening independent, given the assumption of a
reddening law. Moreover, \citet{marconi15} showed that the ($V$, $B-V$)
Wesenheit relation presents the remarkable property of also being independent of
metallicity, over a wide range.  This implies that a possible intrinsic metallicity
dispersion, such as that found in the local group dwarf galaxies Tucana 
\citep{bernard08} and Sculptor \citep{martinezvazquez15}, does not affect 
the distance determination.

\begin{figure}
  \includegraphics[scale=0.6]{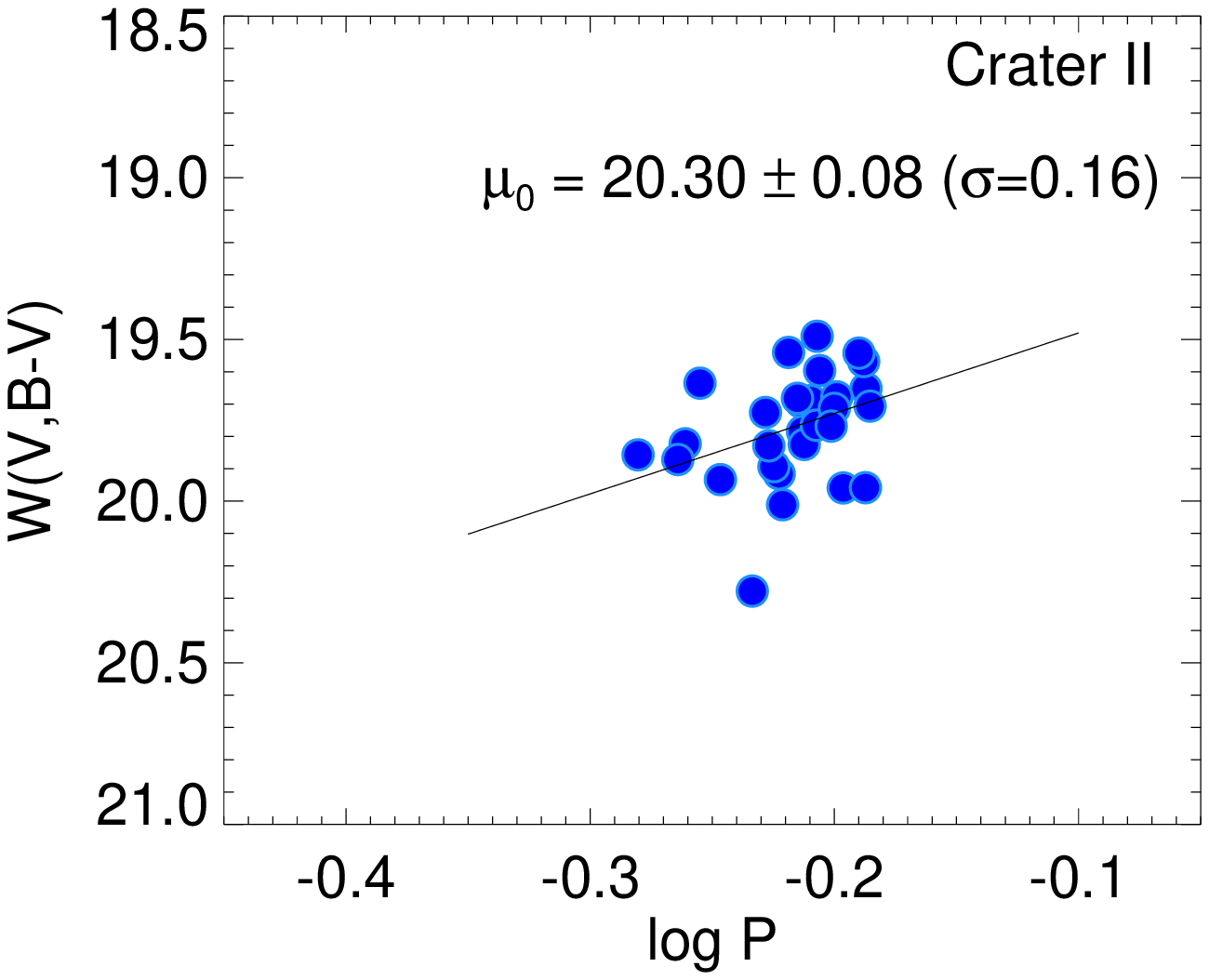}
  \caption{W($V$, $B-V$)-LogP relation for the 24 RRab and the 4 fundamentalized RRc, 
  in Crater II. The black line is the fit of the data to the theoretical slope predicted by
  \citep{marconi15}. The zero-point of this relation, give us the first estimation of the 
  distance modulus ($\mu_0$) for Crater II based on a standard candle.}
     \label{fig:pwr}
  \end{figure}

We adopted the ($V$, $B-V$) Wesenheit relation from \citet{marconi15} and derived the
distance modulus to Crater~II, obtaining (m-M)$_0$ = 20.30$\pm$0.08 mag 
($\sigma$=0.16 mag). This is illustrated in Figure \ref{fig:pwr}, which shows the
W($V$, $B-V$) magnitude as a function of the logarithm of the period for 28 RRL stars.
The peculiar, two uncertain, and the three Blazkho stars have been excluded, and the period 
of the remaining four RR$c$ type have been fundamentalized. The result is in 
good agreement with the value provided by \citet{torrealba16a}, (m-M) = 20.35, 
based on the comparison with theoretical isochrones and with the distance
estimated by \citet{joo18} based on a larger sample of RRL stars, (m-M)$_0$ = 20.25$\pm$0.1.

%%%%%%%%%%%%%%%%%%%%%%%%%%%%%%%%%%%%%%%%%%%%%%%%%%%  FIG 7   %%%%%%%%%%%%%%%%%%%%%%%%%%%%%%%%%%%%%%%%%%5
%%%%%%%%%%%%%%%%%%%%%%%%%%%%%%%%%%%%%%%%%%%%%%%%%%%%%%%%%%%%%%%%%%%%%%%%%%%%%%%%%%%%%%%%%%%%%%%%55 

\section{Metallicity distribution}\label{sec:metal}
	
\begin{figure*}
  \includegraphics[scale=0.8]{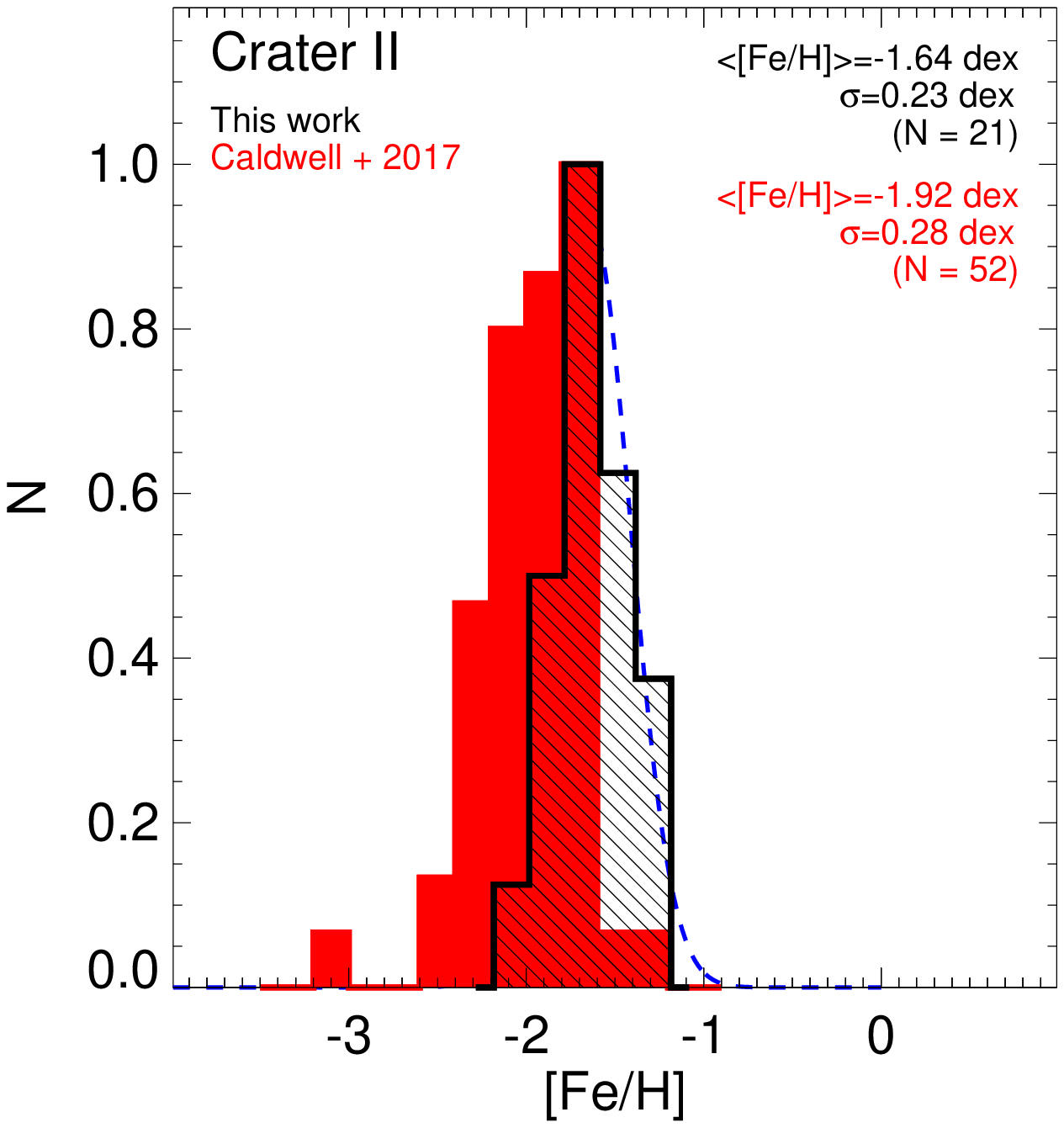}
  \caption{Metallicity distribution of 19 RRL stars in Crater~II as derived from the 
  theoretical $I$-PLZ relation presented in \citep{marconi15}. The red histogram shows 52
  the metallicity derived spectroscopically by \citet{caldwell17a}.}
     \label{fig:metallicity}
  \end{figure*}

\citet{martinezvazquez16a} and \citet{braga16} introduced the use of the $I$-PLZ
relation as a metallicity rather than a distance indicator. If the
distance is provided by an independent indicator the individual metallicity can 
be calculated for each RRL star, given its absolute M$_I$
magnitude and period. In the case of the current data set, we can take
advantage of the three filters available: $B$, $V$, and $I$. First we adopted
the metal-independent distance estimate based on the W($V$, $B-V$) Wesenheit function
derived in \S \ref{sec:distance}. This was used to derive an absolute $I$ magnitude
for each RRL star. Second, we applied the inverse of the $I$-PLZ relation to derive 
individual metallicity (rather than individual distances as typically done)
The derived metallicty distribution for RRL stars is shown as a normalized
dashed histogram in Figure \ref{fig:metallicity}.

%%%%%%%%%%%%%%%%%%%%%%%%%%%%%%%%%%%%%%%%%%%%%%%%%%%  FIG 8 %%%%%%%%%%%%%%%%%%%%%%%%%%%%%%%%%%%%%%%%%%5
%%%%%%%%%%%%%%%%%%%%%%%%%%%%%%%%%%%%%%%%%%%%%%%%%%%%%%%%%%%%%%%%%%%%%%%%%%%%%%%%%%%%%%%%%%%%%%%%55 

We derived individual metallicities for a sample of 21 RR$ab$ stars with good
light curves. We did not include first overtone pulsators, the peculiar star,
the Blazhko ones, the two candidates and four stars with poorly constrained 
light curve in the $I$ band (V010, V015, V022, V023). The distribution cover
a range between [Fe/H] $\sim$ --2.2 and [Fe/H] $\sim$ --1.2
dex, and appears symmetric around a well defined peak. A Gaussian fit to the
distribution provides a mean metallicity [Fe/H]= --1.64 dex, with a dispersion
$\sigma$=0.21 dex. \citet{martinezvazquez16a} discussed the resolution of the method
by analysing the population of RRL stars of globular clusters with no 
spectroscopically measured
intrinsic metallicity spread. The metallicity distribution derived for
22 RRL stars in the LMC cluster Reticulum provided a dispersion $\sigma$=0.25
dex. This suggests that the intrinsic metallicity dispersion of RRL stars in
Crater II is compatible with being basically null. It is also worth stressing 
that the adopted distance affects the metallicity estimate as a zero point: longer
distances produce a more metal-poor distribution (by 0.6 dex for a 0.1 mag shift).
However, the distribution shape, hence its dispersion, is unaffected.

The derived mean metallicity is in good agreement with that found by \citet{joo18},
$<$[Fe/H]$>$= -1.65$\pm$0.15 on the basis of the period-amplitude-metallicity relation 
by \citet{alcock00}, though our dispersion ($\sigma$=0.21 dex) is smaller than
their value ($\sigma$=0.31 dex).
The red histogram in the Figure  \ref{fig:metallicity} shows the metallicity distribution of the
62 stars identified by \citet{caldwell17a} as bona-fide members of Crater\,II 
on the basis of their radial velocity and proximity to the nominal Crater~II RGB.
Metallicity estimates have been homogeneized to the same scale, correcting for
the different solar abundances adopted (log([Fe/H])/log([Fe/H])$_{\odot}$=7.50). The comparison 
discloses substantial agreement between the two, completely independent, estimates.
A gaussian fit provides similar dispersion ($\sigma$=0.28 vs 0.21), with the value
from RGB stars slightly smaller than for the RRL stars. The small offset between the
two distributions can be explained by the 0.05 mag difference in the distance
adopted (20.35 vs 20.30 mag), and it is well within the uncertainties. Apparently
the spectroscopic sample presents a more extended tail to the metal-poor side.
We verified that this is not due a variation in the characteristics of
the stellar populations with radius, since the \citet{caldwell17a} sample 
covers a wider area of Crater~II. The difference is possibly due to a random 
fluctuation due to the relatively small sample of RRL stars used.

Overall, the metallicity distribution of a purely old sample of RRL stars is similar
to that of RGB stars, which in principle can include populations
younger and/or too metal-rich to have counterparts in the RRL stars population. This
suggests that Crater~II experienced a quick (within the formation time
scale of RRL stars progenitors) and limited chemical evolution.
This is supported by the lack of High-Amplitude Short-Period (HASP) RRL stars
\citep{fiorentino15a}.
These stars are a strong indication that their progenitors
belong to a population at least as metal-rich as [Fe/H]=--1.5 dex.
The Bailey diagram of Crater~II is devoid of such stars
(P$_{ab,min}$=0.566652d), suggesting that the upper limit to the 
metallicity distribution of RRL stars is significantly lower than [Fe/H]=--1.5 dex
\citep{fiorentino17}.

\section{Discussion and final remarks}\label{sec:discussion}

Crater~II is a dwarf galaxy with many fascinating properties.
The galaxy size-magnitude plot (e.g., \citealt{torrealba16a}) shows no other MW
satellite with similar properties. It is the fourth most physically extended MW
satellite ($r_h \approx$1 kpc) (after the Magellanic Clouds and Sagittarius)  
{\it and} one of the galaxies with the lowest known surface brightness ($\mu_{V} \sim$ 
31.2 mag). Only Andromeda XIX lies in a similar position in this plot. Therefore, Crater~II
appears as either over-large or under-luminous, compared to most other LG galaxies. 

Analyzing INT/WFC data we have found a substantial population of RRL stars in 
the central regions of Crater~II. The RRL stars that we have discovered lack any central
concentration and are widely distributed over our whole field. Thus, our 
sample of variable stars is likely incomplete and indicates that the galaxy 
extends outside the surveyed field of view. This was already expected, given
the stellar distribution in the discovery paper \citep{torrealba16a}. 

Interestingly, our search for variable stars did not reveal any Anomalous
Cepheids (AC) in the central regions of Crater~II. Assuming the empirical relation found by 
\citet{mateo95} between the fraction of ACs and the luminosity of the host galaxy,
one would expect very few of them, at most 1--2 for M$_V$=--8 mag \citep{torrealba16a}.
\citet{fiorentino12b} argued that this relation may be used to identify
dwarf galaxies with an important fraction of intermediate-age population, as in
this case the number of ACs increases with respect to a purely old sample
descending from primordial binary stars. In this sense, the absence of ACs in
Crater~II can be compared to the sizable samples detected in galaxies of similar
luminosity, such as And XIX, which hosts 8 AC \citep{cusano13}, possibly indicating
a significant intermediate age population, and Leo~T \citep{clementini12}, where 
indeed a strong intermediate-age population is present. This may be 
considered indirect evidence that Crater~II has a negligible, if any, younger 
population. 

Comparison with theoretical isochrones supports this conclusion and suggests
that the bulk of star formation in Crater~II occurred at an early epoch. The presence
of a sequence of objects bluer and brighter than the TO of the old population is 
compatible with a population of BSS. Nevertheless, deeper data is necessary in order
to set stronger constraints on the quenching epoch and on the existence of a minority
intermediate-age population.

The metallicity derived from a sample of 21 RRL stars reveals a narrow 
distribution peaked at [Fe/H]=--1.64, with an observed dispersion compatible
with a negligible intrinsic dispersion. The chemical properties of RRL stars
are in agreement with those of the sample of 62 RGB stars investigated
by \citet{caldwell17a}, both in terms of mean metallicity and dispersion.
Since the RGB can be populated by a mix of populations with a significantly
broader range of age and metallicity than a purely old sample of RRL stars, this
suggests that Crater~II does not host any strong population significantly
more metal-rich than the RRL stars. Similarly, the narrow metallicity 
distribution may suggest the absence of a significant tail of very low 
metallicity stars such as are found in other dwarf galaxies of similar brightness, 
like Eridanus II \citep{li17} or CVnI \citep{kirby08}.

Finally, a striking feature in the CMD of Crater~II is the morphology 
of the HB. In fact, while the red HB is well populated, only a handful of 
objects are located in the region of the blue HB, for stars
cooler than 0.0 $< B - I <$ 0.5~mag. 
This is a strong peculiarity compared to bright MW satellites 
which {\it always\/} present a sizable sample of RRL stars together with
a well populated HB for colours bluer than the instability strip. This is
true independently of the details of the early SFH. This occurrence is 
present in the most metal-poor systems such as Ursa Minor (which has the
bluest HB), but also in those with strong intermediate-age component such
as Carina, Fornax or Leo~I. Despite the redder, on average, HB morphology
\citep{dacosta96,dacosta00,dacosta02}, such an extreme case is not common 
among the M31 satellites either. The analysis of twenty galaxies by
\citet{martin07} suggests that only And XXIV may lack blue HB
stars. Interestingly, the luminosity of And XXIV and the metallicity ([Fe/H]=--1.8$\pm$0.2 
\citealt{richardson11}) are similar to that of 
Crater~II (M$_V$=--8.5 mag), but the spatial extent of And XXIV is smaller (r$_h$=680 kpc). However, no search for variable stars has been performed in
And XXIV, and the data of \citet{martin17} cover about 36\% of its
extent. A deeper analysis of the SFH and the stellar content of 
this galaxy would be of extreme interest to compare with Crater~II.

The HB morphology of Crater~II becomes even more peculiar when compared with  HB
morphologies of Galactic Globular Clusters (GCs). Indeed, GCs with  metal abundances
in the range -1.9 $<$ [Fe/H] $<$ -1.5 present a predominantly blue HB  (e.g.
NGC~6535, NGC~6144, NGC~6541), or a redder morphology but still with a well
populated blue part (e.g. NGC~7006, NGC~5272). The few clusters with entirely red
HBs do not host any RRL (e.g. AM 1, Palomar 14, Pyxis), unlike the case for
Crater~II. Possibly, the only GC with HB resembling that of Crater~II  is the
peculiar Ruprecht 106 \citep{dotter11,kaluzny95}. This is one of the very few GCs
which does not display chemical abundance variations typical of multiple populations
\citep{villanova13,bastian17}. Moreover and even more importantly, Ruprecht 106 has
also the remarkable feature  of not presenting enhancement in the $\alpha$-elements
\citep{villanova13}.  This means that the empirical scenario becomes even more
puzzling, since  dwarf galaxies in the metal-poor regime ([Fe/H]$\le$ -1.5) display
a well  defined $\alpha$-enhancement \citep{cohen09b,hendricks14a,fabrizio15}.

The current empirical evidence indicates that the HB morphology of Crater~II  is too
red for its metallicity. The paucity of blue HB stars, once confirmed,  is a real
conundrum, since theory and observations are suggesting that  the metal content is
the main parameter in driving the HB morphology  \citep{salaris05}.  The lack of blue HB stars is an additional indication that the
very metal-poor (Z $<$ 0.0003) component of Crater~II is small, if present.  We are
therefore presented with a i) peculiarly very extended, relatively  faint and low
surface brightness galaxy with, however, ii) not extremely low mean metallicity, but
low metallicity dispersion and possibly lacking a tail toward extremely
metal-poor stars typical of galaxies of similar brightness. 

What mechanisms may have originated a dwarf with a peculiar HB morphology, a 
low metallicity dispersion and possibly lacking of extremely metal-poor stars? 

\citet{caldwell17a} found a far too low central velocity dispersion ($\sigma_{v}$=2.7 Km s$^{-1}$), 
difficult to explain within the standard $\Lambda$-Cold Dark Matter scenario given 
the galaxy's observed large size and moderate luminosity. To overcome this, it has been
suggested that strong mass loss ($\sim$90\%) must have occurred in Crater~II
\citep{sanders18,fattahi18}. This prediction appears in good agreement with
the first orbit determination based on Gaia DR2 \citep{fritz18a}, which supports
that Crater~II is on a relatively eccentric orbit with pericentre close to 20~kpc.
As a result, Crater~II may have already experienced a few passages through denser
MW regions. The earliest pericentric passage(s) may have stripped the oldest, 
most metal poor component, while the galaxy must have been massive enough to 
avoid total tidal disruption and keep sufficient mass to undergo further star 
formation which would have given rise to the moderately metal rich population 
currently seen in the center of the galaxy where we have sampled (within 
half its $r_h$). This scenario is similar to the one that can be inferred for 
the Sagittarius dwarf spheroidal galaxy based on the stellar composition and 
SFHs at different positions along its tidal stream \citep{martinezdelgado04, deboer15}. 

However, population gradients are commonly observed in LG dwarf galaxies 
\citep{harbeck01,tolstoy04,bernard08,martinezvazquez15,martinezvazquez16a}, 
in the sense that the younger/more metal-rich components
are more centrally concentrated (and this is reflected in the HB getting bluer 
when moving to the external regions). Additionally, deeper data will 
allow us to put more stringent limits on the possible duration of the star formation epoch. 
These new data are thus necessary to put stronger constraints on the properties of 
the old populations of this galaxy and their possible radial gradients. The existence 
(or not) of a very metal poor population at larger galactocentric radius will help 
constrain its early evolution, the possible effects of early stripping episodes, 
and therefore even its history of interactions with the MW, and thus possibly its 
orbit. The stellar distribution in a large area may also reveal the presence of 
tidal tails, which however may be challenging to detect due to the extremely low 
surface brightness.

\section*{Acknowledgements}

This research has been supported by the Spanish Ministry of Economy and 
Competitiveness (MINECO) under the grant AYA2014-56795-P. This research has made 
use of the NASA/IPAC Extragalactic Database (NED) which is operated by the Jet 
Propulsion Laboratory, California Institute of Technology, under contract with 
the National Aeronautics and Space Administration.

%%%%%%%%%%%%%%%%%%%%%%%%%%%%%%%%%%%%%%%%%%%%%%%%%%

%%%%%%%%%%%%%%%%% APPENDICES %%%%%%%%%%%%%%%%%%%%%

%%%%%%%%%%%%%%%%%%%%%%%%%%%%%%%%%%%%%%%%%%%%%%%%%%

\appendix

\section{Online material}\label{sec:appendix}

Time series photometry for detected variable stars in the central regions of Crater~II
is listed in Table \ref{tab:photometry}. The table presents, for each star, the 
modified Heliocentric Julian Date of mid-exposure (HJD - 2,400,000, columns 1, 4, 7), 
individual $B$, $V$, and $I$ measurements (columns 2, 5, 8) and their uncertainties (columns
3, 6, 9). The full version of the table is available on the online version of the paper.

\begin{table*}
\begin{scriptsize}
\centering
\caption{Photometry of the variable stars in Crater II dSph.} 
\label{tab:photometry}
%\hspace{+0.5cm}
  \begin{tabular}{ccccccccc} 
 \hline
MHJD$^*$    &    $B$    &    $\sigma_B$    &    MHJD$^*$    &    $V$    &    $\sigma_V$    &    MHJD$^*$    &    $I$    &    $\sigma_I$   \\
\hline
\multicolumn{9}{c}{CraterII-V001} \\
\hline
   57538.3730  &   20.862  &    0.192  &   57538.4276  &   20.737  &    0.023  &   57538.4312  &   20.069  &    0.039 \\
   57538.4249  &   21.214  &    0.028  &   57540.4067  &   20.227  &    0.016  &   57540.4102  &   19.839  &    0.028 \\
   57539.4280  &   21.318  &    0.036  &   57540.4395  &   20.306  &    0.014  &   57540.4137  &   19.769  &    0.030 \\
   57540.4039  &   20.486  &    0.018  &   57541.4030  &   20.631  &    0.021  &   57540.4430  &   19.812  &    0.024 \\
   57540.4368  &   20.606  &    0.021  &   57541.4296  &   20.647  &    0.018  &   57541.4077  &   19.998  &    0.024 \\
   57541.4002  &   21.062  &    0.023  &   57847.4935  &   20.620  &    0.026  &   57541.4269  &   20.003  &    0.027 \\
   57541.4324  &   21.147  &    0.024  &   57847.5050  &   20.650  &    0.025  &   57781.6373  &   19.832  &    0.026 \\
   57781.6234  &   20.647  &    0.016  &   57847.5247  &   20.675  &    0.027  &   57781.6801  &   19.797  &    0.016 \\
   57781.6678  &   20.792  &    0.012  &   57847.5465  &   20.688  &    0.034  &   57781.6968  &   19.868  &    0.014 \\
   57781.6848  &   20.828  &    0.011  &   57847.5664  &   20.738  &    0.034  &   57781.7140  &   19.840  &    0.015 \\
   57781.7017  &   20.865  &    0.011  &   57847.5873  &   20.712  &    0.028  &   57781.7300  &   19.856  &    0.013 \\
   57781.7190  &   20.893  &    0.013  &   57847.6049  &   20.804  &    0.029  &   57781.7525  &   19.910  &    0.019 \\
   57781.7395  &   20.939  &    0.014  &   57848.4812  &   20.864  &    0.043  &   57782.6171  &   19.942  &    0.032 \\
   57781.7570  &   20.994  &    0.019  &   57848.4992  &   20.761  &    0.042  &   57782.6321  &   19.953  &    0.024 \\
   57782.6106  &   21.170  &    0.022  &   57848.5345  &   20.730  &    0.040  &   57782.6459  &   20.018  &    0.023 \\
   57782.6235  &   21.136  &    0.019  &   57848.5526  &   20.878  &    0.044  &   57782.6573  &   20.073  &    0.021 \\
   57782.6381  &   21.165  &    0.018  &   57848.5747  &   20.867  &    0.047  &   57782.6688  &   20.037  &    0.018 \\
   57782.6502  &   21.204  &    0.016  &   57848.5940  &   20.954  &    0.047  &   57782.6802  &   20.017  &    0.021 \\
   57782.6616  &   21.204  &    0.015  &   57849.4671  &   20.083  &    0.037  &   57782.6917  &   20.085  &    0.027 \\
   57782.6731  &   21.248  &    0.016  &   57849.4869  &   20.101  &    0.035  &   57782.7031  &   20.109  &    0.024 \\
   57782.6845  &   21.232  &    0.015  &   57849.5126  &   20.162  &    0.034  &   57782.7145  &   20.070  &    0.023 \\
   57782.6959  &   21.237  &    0.015  &   57849.5337  &   20.218  &    0.040  &   57782.7260  &   20.099  &    0.020 \\
   57782.7073  &   21.252  &    0.016  &   57849.5537  &   20.201  &    0.042  &   57782.7374  &   20.134  &    0.028 \\
   57782.7188  &   21.286  &    0.016  &   57849.5724  &   20.355  &    0.051  &   57782.7488  &   20.075  &    0.022 \\
   57782.7303  &   21.280  &    0.015  &   57849.5910  &   20.286  &    0.050  &   57782.7605  &   20.104  &    0.023 \\
   57782.7417  &   21.305  &    0.017  &       ---     &     ---   &    9.999  &   57782.7719  &   20.093  &    0.033 \\
   57782.7533  &   21.300  &    0.018  &       ---     &     ---   &    9.999  &   57800.5651  &   19.951  &    0.041 \\
   57782.7648  &   21.286  &    0.021  &       ---     &     ---   &    9.999  &   57800.5723  &   19.844  &    0.035 \\
   57847.4876  &   21.053  &    0.046  &       ---     &     ---   &    9.999  &   57800.5763  &   19.874  &    0.028 \\
   57847.4975  &   21.110  &    0.048  &       ---     &     ---   &    9.999  &   57800.5803  &   19.823  &    0.030 \\
   57847.5013  &   21.094  &    0.046  &       ---     &     ---   &    9.999  &   57800.5843  &   19.825  &    0.029 \\
   57847.5208  &   21.178  &    0.043  &       ---     &     ---   &    9.999  &   57800.5884  &   19.729  &    0.026 \\
   57847.5425  &   21.196  &    0.050  &       ---     &     ---   &    9.999  &   57800.5924  &   19.723  &    0.022 \\
   57847.5625  &   21.142  &    0.046  &       ---     &     ---   &    9.999  &   57800.5964  &   19.656  &    0.028 \\
   57847.5833  &   21.249  &    0.049  &       ---     &     ---   &    9.999  &   57800.6004  &   19.701  &    0.029 \\
   57847.6010  &   21.248  &    0.033  &       ---     &     ---   &    9.999  &   57800.6044  &   19.671  &    0.028 \\
   57848.4773  &   21.211  &    0.059  &       ---     &     ---   &    9.999  &   57800.6085  &   19.647  &    0.029 \\
   57848.4952  &   21.181  &    0.072  &       ---     &     ---   &    9.999  &   57800.6141  &   19.635  &    0.025 \\
   57848.5130  &   21.173  &    0.071  &       ---     &     ---   &    9.999  &   57800.6182  &   19.645  &    0.028 \\
   57848.5305  &   21.084  &    0.060  &       ---     &     ---   &    9.999  &   57800.6222  &   19.666  &    0.028 \\
   57848.5487  &   21.262  &    0.064  &       ---     &     ---   &    9.999  &   57800.6262  &   19.715  &    0.031 \\
   57848.5707  &   21.373  &    0.072  &       ---     &     ---   &    9.999  &   57800.6302  &   19.693  &    0.029 \\
   57848.5900  &   21.339  &    0.069  &       ---     &     ---   &    9.999  &   57800.6342  &   19.693  &    0.031 \\
   57849.4631  &   20.306  &    0.048  &       ---     &     ---   &    9.999  &   57800.6382  &   19.666  &    0.029 \\
   57849.4830  &   20.259  &    0.041  &       ---     &     ---   &    9.999  &   57800.6422  &   19.685  &    0.031 \\
   57849.5086  &   20.284  &    0.051  &       ---     &     ---   &    9.999  &   57800.6462  &   19.689  &    0.029 \\
   57849.5298  &   20.473  &    0.057  &       ---     &     ---   &    9.999  &   57800.6503  &   19.692  &    0.031 \\
   57849.5497  &   20.509  &    0.063  &       ---     &     ---   &    9.999  &   57800.6543  &   19.712  &    0.030 \\
   57849.5685  &   20.507  &    0.060  &       ---     &     ---   &    9.999  &   57800.6583  &   19.717  &    0.029 \\
   57849.5870  &   20.657  &    0.073  &       ---     &     ---   &    9.999  &   57800.6623  &   19.727  &    0.032 \\
       ---     &     ---   &    9.999  &       ---     &     ---   &    9.999  &   57800.6663  &   19.752  &    0.032 \\
       ---     &     ---   &    9.999  &       ---     &     ---   &    9.999  &   57800.6703  &   19.700  &    0.032 \\
       ---     &     ---   &    9.999  &       ---     &     ---   &    9.999  &   57800.6744  &   19.653  &    0.031 \\
       ---     &     ---   &    9.999  &       ---     &     ---   &    9.999  &   57800.6784  &   19.804  &    0.032 \\
       ---     &     ---   &    9.999  &       ---     &     ---   &    9.999  &   57800.6824  &   19.782  &    0.027 \\
       ---     &     ---   &    9.999  &       ---     &     ---   &    9.999  &   57800.6864  &   19.715  &    0.033 \\
       ---     &     ---   &    9.999  &       ---     &     ---   &    9.999  &   57800.6904  &   19.755  &    0.036 \\
       ---     &     ---   &    9.999  &       ---     &     ---   &    9.999  &   57800.6945  &   19.759  &    0.034 \\
       ---     &     ---   &    9.999  &       ---     &     ---   &    9.999  &   57800.6985  &   19.744  &    0.030 \\
       ---     &     ---   &    9.999  &       ---     &     ---   &    9.999  &   57800.7026  &   19.819  &    0.040 \\
       ---     &     ---   &    9.999  &       ---     &     ---   &    9.999  &   57800.7066  &   19.760  &    0.034 \\
       ---     &     ---   &    9.999  &       ---     &     ---   &    9.999  &   57800.7106  &   19.756  &    0.036 \\
       ---     &     ---   &    9.999  &       ---     &     ---   &    9.999  &   57800.7146  &   19.714  &    0.035 \\
       ---     &     ---   &    9.999  &       ---     &     ---   &    9.999  &   57800.7186  &   19.758  &    0.035 \\
\hline
\end{tabular}
\end{scriptsize}
\end{table*}

% Don't change these lines
\bsp	% typesetting comment
\label{lastpage}
\end{document}